\documentclass[a4paper]{article}

\usepackage{INTERSPEECH2022}
\usepackage{graphicx}
\usepackage{color}
\usepackage{multirow}
\usepackage{etoolbox,siunitx}
\usepackage{hyperref}
\usepackage{pifont}

\newcommand{\xmark}{\ding{55}}%

\title{End-to-End Integration of Speech Recognition, Speech Enhancement, and Self-Supervised Learning Representation}

\name{Xuankai Chang$^1$, Takashi Maekaku$^2$, Yuya Fujita$^2$, Shinji Watanabe$^1$}
\address{
  $^1$Carnegie Mellon University, PA, USA \\
  $^2$Yahoo Japan Corporation, Tokyo, JAPAN
}
\email{\{xuankaic,swatanab\}@andrew.cmu.edu, \{tmaekaku,yuyfujit\}@yahoo-corp.jp}

\begin{document}

\ninept
\maketitle
\begin{abstract}
  This work presents our end-to-end (E2E) automatic speech recognition (ASR) model targetting at robust speech recognition, called Integraded speech Recognition with enhanced speech Input for Self-supervised learning representation (IRIS). Compared with conventional E2E ASR models, the proposed E2E model integrates two important modules including a speech enhancement (SE) module and a self-supervised learning representation (SSLR) module. The SE module enhances the noisy speech. Then the SSLR module extracts features from enhanced speech to be used for speech recognition (ASR). To train the proposed model, we establish an efficient learning scheme. Evaluation results on the monaural CHiME-4 task show that the IRIS model achieves the best performance reported in the literature for the single-channel CHiME-4 benchmark ($2.0\%$ for the real development and $3.9\%$ for the real test) thanks to the powerful pre-trained SSLR module and the fine-tuned SE module.
\end{abstract}
\noindent\textbf{Index Terms}: robust automatic speech recognition, self-supervised learning, speech enhancement, deep learning

\section{Introduction}
\label{sec:intro}
In the past decade, deep learning has significantly pushed the development of automatic speech recognition (ASR) moving forward. Many interesting models and technologies have been proposed. Deep neural network-hidden Markov model (DNN-HMM) based hybrid system \cite{hinton2012deep} is one of the them. DNN-HMM hybrid ASR systems usually train a DNN to predict frame-aligned states, e.g. context-dependent phonemes. Recently, end-to-end speech recognition systems have become more and more popular. Several end-to-end ASR technologies were proposed, including connectionist temporal classification (CTC) \cite{graves2006connectionist}, Transducer \cite{graves2013speech} and attention-based encoder-decoder \cite{chan2015listen,kim2017joint,watanabe2018espnet}. A lot of existing speech recognition techniques exhibit strong performance in clean conditions. However, applying speech recognition in noisy environments is still challenging, especially in the monaural case. DNN-HMM hybrid ASR systems still outperform E2E ASR system on a well-known noisy speech corpus\cite{yang2022conformer}, CHiME-4 corpus \cite{vincent2017analysis}.

Usually, speech signals recorded in the real scenarios contain unpredicted noise. The noise is from the environment or the device imperfections, which degrades the ASR performance. The existing solutions to address the noisy speech recognition can be summarized as two categories. One is to train the ASR model robust to noise \cite{hannun2014deep, shinohara2016adversarial,kim2017efficient}. The other is to use an dedicated model to improve the intelligibility of the noisy speech before sending it to the ASR model. Such preprocessing is one of the important topics in speech research, called speech enhancement (SE) or denoising \cite{loizou2007speech}. The SE model and the ASR model can be trained separately or jointly  \cite{ochiai2017multichannel,narayanan2014joint,subramanian2019speech}. However, it is well known that the monaural speech enhancement techniques produce distortions which deteriorate the ASR performance \cite{iwamoto2022bad,Closing-Zhang2021}.

Recently, self-supervised learning representations (SSLR) have demonstrated great potential in improving the speech recognition \cite{baevski2020wav2vec,zhang2020pushing,hsu2021hubert,chang2021exploration}. One primary drawback of current SSLR models is that the pre-training cost is too high for most of the research groups. As an alternative solution, some researchers fine-tune the pre-trained SSLR models to get their customized version \cite{pasad2021layer}. In our previous study \cite{chang2021exploration}, we have shown that directly using the pre-trained Wav2Vec2.0 \cite{baevski2020wav2vec} and HuBERT \cite{hsu2021hubert} for feature extraction improves the ASR performance. However, the improvement on mismatched conditions is usually limited. The result of CHiME-4 corpus in \cite{chang2021exploration} shows the word error rate (WER) reduction for the multi-channel data with beamforming are much better than those for the isolated single channel. Because the audio of the latter set is relatively noisier than the former one. We believe that it is due to the mismatch between the pre-training and the target task. Wav2Vec2.0 and HuBERT models were pre-trained on the LibriLight \cite{librilight} data, a clean read English speech corpus. Later, WavLM \cite{chen2021wavlm} was proposed to learn a representation model on simulated noisy / overlapped speech. In another recent work, Wang \textit{et al.}\cite{wang2021wav2vec} proposed a noisy robust SSLR model based on Wav2Vec2.0, which also shows promising results on CHiME-4. But the model is not publicly available.

In this work, we propose a new model, called IRIS, for robust speech recognition, which integrates an SE module, an SSLR module and an ASR module into a single end-to-end model. We extensively investigate the benefits of the SE module and the SSLR module for robust speech recognition. Through experiments, we establish an efficient training scheme for the proposed E2E IRIS model. Finally, we show that our model achieves state-of-the-art performance on the single-channel CHiME-4 ASR tasks.

\section{IRIS Model}
\label{sec:system}
\begin{figure}[!tpb]
    \centering
    \includegraphics[width=\linewidth]{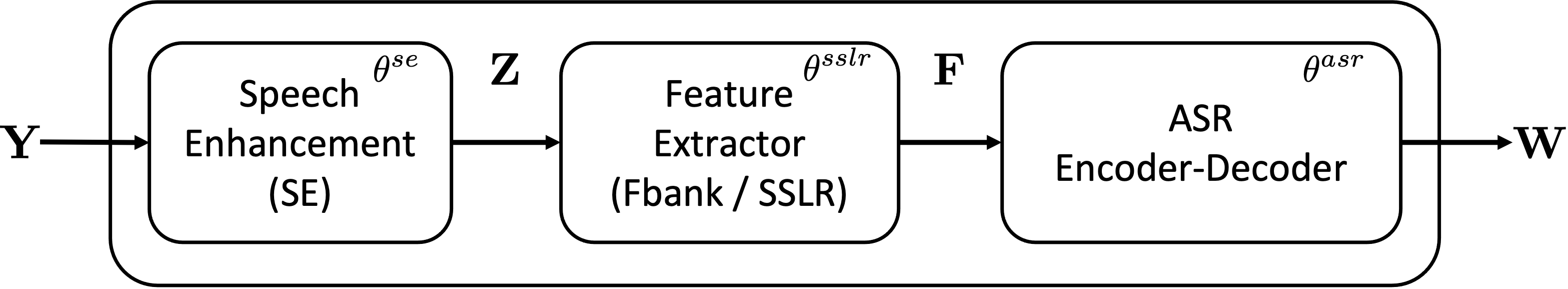}
    \caption{Overview of the proposed end-to-end model.}
    \label{fig:model}
    \vspace{-5pt}
\end{figure}

We describe the proposed IRIS model in this section. The model includes a speech enhancement module (SE) and an self-supervised learning representation (SSLR)-based ASR (SSLR-ASR) module, shown in Figure~\ref{fig:model}. Each module can be trained separately. Then whole model can be fine-tuned with the objectives of speech enhancement and recognition. For the convenience of the following discussions, we denote the noisy speech input as $\mathbf{Y}$. 

\subsection{Speech Enhancement}
\label{ssec:se}

Most of the data collected in real scenarios contains not only speech signal but also undesired noise and reverberation. The target of speech enhancement is to keep the speech signal from data and to suppress the undesired signals. We denote the SE process as the following:
\begin{align}
    \mathbf{Z} &= \text{SE}(\mathbf{Y}; \,\theta^{\text{se}}),
\end{align}
where $\mathbf{Z}$ is the enhanced speech and $\theta^{\text{se}}$ represents the parameters of the SE model.

A lot of powerful speech enhancement techniques have been proposed. In this work, we choose the Conv-TasNet proposed in \cite{luo2019conv} as the SE module. Conv-TasNet is a very successful model for end-to-end time-domain speech enhancement. Besides this, many other strong end-to-end time-domain speech enhancement models were proposed before \cite{luo2018tasnet,pandey2019tcnn,luo2020dual}. The advantage of the time-domain speech enhancement model is that we do not need to care about the phase when we generate enhanced signals. This might be helpful to reduce the distortion generated by speech enhancement models. Without loss of generality, any speech enhancement models can be used in our model.

\subsection{SSLR-ASR}
\label{ssec:sslr-asr}

\subsubsection{E2E-ASR}
\label{sssec:e2e-asr}

To recognize the speech, we use an E2E-ASR model. If we denote the input speech signal as $\mathbf{Z}$, the feature of the speech as $\mathbf{F}$ and the text as $\mathbf{W}$, we can write the ASR process as:
\begin{align}
    \mathbf{F} &= \text{FeatureExtraction}(\mathbf{Z}), \label{eq:feature} \\
    \mathbf{W} &= \text{ASR}(\mathbf{F}; \,\theta^{\text{asr}}),
\end{align}
where $\theta^{\text{asr}}$ represents the parameters of the ASR model. In this work, we use the joint CTC / attention-based encoder-decoder framework proposed in \cite{kim2017joint} to build our E2E-ASR model. More details can be referred to \cite{kim2017joint,watanabe2018espnet}. It is worth to note that the choice of ASR is not limited to a specific architecture.

\subsubsection{SSLR}
\label{sssec:sslr}
Conventional ASR models use energy-based features such as log Mel-Filterbanks (Fbank) and mel-frequency cepstral coefficients (MFCC). In our previous work \cite{chang2021exploration}, we have shown that replacing the energy-based features with SSLRs can improving the performance of E2E-ASR. In this way, the Eq.~\ref{eq:feature} would be rewritten as:
\begin{align}
    \mathbf{F} &= \text{SSLR}(\mathbf{Z}; \,\theta^{\text{sslr}}), \label{eq:sslr} 
\end{align}
where $\theta^{\text{sslr}}$ represents the parameters of the SSLR model.

SSLR models are learning-based speech representations. SSLR models are trained using large amount of unlabelled data. In this study, we propose to use WavLM proposed in \cite{chen2021wavlm} to improve robust speech recognition. Similar to HuBERT \cite{hsu2021hubert}, WavLM is trained to predict pseudo-labels of masked segments. In this way, WavLM / HuBERT learns the linguistic information from speech. Moreover, the WavLM learns to handle the noise from the speaker identification, separation, and diarization tasks. The training data of WavLM includes 60k hours of Libri-Light \cite{librilight}, 10k hours of GigaSpeech \cite{chen2021gigaspeech}, and 24k hours of VoxPopuli \cite{wang2021voxpopuli}. This motivates us to use WavLM to extract features for noisy speech.

\subsection{End-to-End IRIS Model}
\label{ssec:joint}

Although WavLM shows good performance on the noisy speech input in downstream tasks, such as speaker identification, separation and diarization tasks \cite{chen2021wavlm}, it is still a question whether it can handle various noises. We propose to use a speech enhancement model to help the WavLM. Our end-to-end model can be written as:
\begin{align}
    \mathbf{W} &= \text{ASR}\ (\ \text{SSLR}\ (\ \text{SE}(\mathbf{Y}; \, \theta^{\text{se}}); \,\theta^{\text{sslr}}\ ); \, \theta^{\text{asr}}\ )
\end{align}

The proposed model adopts a modularized design, where the SE module enhances the input noisy speech, SSLR module extracts the feature and the ASR mudule generates the transcription. We directly use the pre-trained SSLR models from existing works, which are publicly available. Usually, SSLR models are very large, which makes it difficult to train the whole model. To address this issue, all three modules are initialized by pre-trained models with parameters $\hat{\theta}^{\text{se}}$, $\hat{\theta}^{\text{sslr}}$ and $\hat{\theta}^{\text{asr}}$, respectively. Then the parameters of SE and ASR are fine-tuned to be get better performance.

\section{Experimental Setup}
\label{sec:setup}

\subsection{CHiME-4 Corpus}
\label{ssec:corpus}

We carried out all the experiments on the CHiME-4 corpus \cite{vincent2017analysis}. The dataset contains real and simulated six-channel noisy recordings of speech from Wall Street Journal (WSJ0) corpus. The recordings cover four noisy scenarios including bus, cafe, pedestrian and street. There are 1,600 real and 7,138 simulated utterances for training, 1,640 real and 1,640 simulated utterances for development, and 1,320 real and 1,320 simulated utterances for test.

To train the ASR model, all six channels except the second channel of the CHiME-4 training set recordings are used. This brings slight improvement because the second channel faces backward. Besides the noisy utterances in CHiME-4, the clean Wall Street Journal (WSJ0 + WSJ1) utterances are also used to train the E2E ASR model, based on the original ESPnet CHiME-4 recipe\footnote{\url{https://github.com/espnet/espnet/tree/master/egs2/chime4/asr1}}. To train the SE model, only the simulated utterances are used. During evaluation, the single-channel development and test sets are used.

\subsection{Configurations}
\label{ssec:config}

In this work, all the models were trained using the ESPnet toolkit \cite{watanabe2018espnet}. We plan to make the configurations, scripts and trained models to be publicly available.

We use a relatively small Conv-TasNet enhancement model to save the computation. The encoder consists of an 1-D convolution layer, with 256 output channel (N). The kernel and stride sizes are 40 and 20 respectively. The decoder has a reverse 1-D convolution layer with corresponding hyper-parameters of encoder. In the separation part, the temporal convolutional network (TCN), 4 convolutional blocks (X) are repeated twice (R). The number of channels (H) and the kernel size (P) in convolutional blocks are 512 and 3, respectively. The bottleneck has 256 channels (B). More detailed meaning of the hyper-parameters can be referred to \cite{luo2019conv}. SI-SNR\cite{luo2018tasnet} is used to computed the enhancement loss between the reference signal and the enhanced signal. The enhancement model is optimized by adam algorithm with learning rate at $1 \times 10^{-3}$.

For the ASR model, we use Transformer block to build the encoder and the decoder. The ASR model contains $12$ encoder and $6$ decoder Transformer layers. For each Transformer layer, the number of attention heads is $4$. The dimension of the linear projection is $2,048$. The encoder uses two convolutional layers to downsample the input feature sequence and the total frame shift is $40 \text{ms}$. The dropout is set to be $0.1$. In addition to the log Mel-Filterbank (Fbank), we use two SSLR models as feature extractor. One is the HuBERT large model trained on libriLight 60k data and the other is the WavLM large model. When using the SSLR as feature extractor, the feature dimension is reduced from $1,024$ to $128$ with a linear layer before input to the encoder.
The ASR model is optimized by adam algorithm with peak learning rate at $1 \times 10^{-3}$ and $20\mathrm{k}$ steps to warm up. Specaug \cite{park2019specaugment} is used for both Fbank and SSLR feature during training. In decoding, we use a Transformer language model based on character level, with weight 1.0 during beam search.

In the proposed IRIS model, each of the modules is initialized by pre-training. SE and ASR parameters are from the pre-trained models described above. The IRIS model is fine-tuned with 10 epochs using both the enhancement and ASR losses. The same optimizer algorithm for ASR model training is used, with learning rate at $5\times 10^{-4}$. During the training of both ASR and IRIS models, the parameters of the SSLR models are not updated.

Unless otherwise mentioned, model averaging is performed over the 10 checkpoints with best accuracy during decoding.

\section{Results}
\label{sec:eval}

We present our experimental results and analysis in this section.

\vspace{-3pt}
\subsection{E2E-ASR Model with SSLRs}
\label{ssec:e2e-asr-results}
\vspace{-3pt}

In this part, we show the evaluation results of ASR models on the monaural CHiME-4 corpus. The word error rates (WERs) of both simulated and real speech recordings are computed on the development and the test sets. The results are shown in Table~\ref{tab:e2e-asr-results}. The results of systems 1-4 are from existing research works. Among them, system 4 is based on E2E-ASR. The rest systems are built by hybrid ASR systems. We can observe that the best performance is achieved by the hybrid ASR method. We have trained systems 5-7. In system 5, we use the conventional Fbank feature to train the E2E-ASR model, the performance of which is worse than system 3 by a large gap. In system 6 and 7, we use the HuBERT and WavLM models, which are pre-trained on large amount of unlabelled data, to extract feature. When using HuBERT to generate speech features, there is no consistent or obvious improvement across all the evaluation data. We conjecture that it is because the HuBERT is only pre-trained on the clean speech. This can be inferred from the performance of system 4 and 7. In system 4, the Wav2Vec2.0-based model was trained with noisy speech data, leading to similar performance as system 3. Likewise, system 7 using WavLM for feature extraction also achieves comparable performance with system 3. The WERs of simulated speech are $5.9\%$ and $8.2\%$ on dev and test sets, respectively, and those of real speech are $4.0\%$ and $4.5\%$. Specially, in the test set, the WERs of real recordings is $28\%$ better than the previous best results. From this results, we find that it is important to use noisy data to train the robust speech SSLR. 

\begin{table}[htb]
\centering
\sisetup{table-format=2.1,round-mode=places,round-precision=1,table-number-alignment = center,detect-weight=true}
\caption{Single-channel CHiME-4 ASR performance (\%WER) of the E2E-ASR model and previous studies on monaural dev and test sets. In system 6 and 7, HuBERT and WavLM models are pre-trained with large amount of unlabelled data.}
\resizebox{1.0\linewidth}{!}{
\begin{tabular}{clc|cc|cc} 
\toprule
\multirow{2}{*}{ID} & \multirow{2}{*}{System} & \multirow{2}{*}{Model} & \multicolumn{2}{c|}{Dev. Set} & \multicolumn{2}{c}{Test Set}  \\ 
\cmidrule{4-7}
                    &                          &   &  Simu.         & Real          & Simu.        & Real           \\ 
\midrule
1                   & Kaldi Baseline \cite{chen2018building} & Hybrid    & 6.81          & 5.58          & 12.15        & 11.42          \\
2                   & Du \textit{et al.} \cite{du2016ustc}  & Hybrid   & 6.61          & 4.55          & 11.81        & 9.15           \\
3                   & Yang \textit{et al.} \cite{yang2022conformer} & Hybrid    & \textbf{4.99} & \textbf{3.35} & 8.61         & 6.25       \\     
4                   & Wav2Vec-Switch \cite{wang2021wav2vec} & E2E    & -             & 3.5           & -            & 6.6            \\
\midrule
5                   & E2E Transformer - Fbank & E2E    & 11.32          & 9.43           & 19.67         & 17.99           \\
6                  & E2E Transformer - HuBERT & E2E     &      11.56         &     9.13          &     18.02         &   20.41 \\
7                  & E2E Transformer - WavLM & E2E     & 5.93           & 4.03           & \textbf{8.25} & \textbf{4.47}   \\
\bottomrule
\end{tabular}
}
\label{tab:e2e-asr-results}
\end{table}
\vspace{-8pt}

\vspace{-3pt}
\subsection{IRIS Model}
\label{ssec:e2e-asr}
\vspace{-3pt}

\begin{table}[htb]
\vspace{-5pt}
\centering
\sisetup{table-format=2.1,round-mode=places,round-precision=1,table-number-alignment = center,detect-weight=true}
\caption{Monaural CHiME-4 ASR performance (\%WER) of the IRIS model. Different combinations of fine-tuning SE (FT. SE) and fine-tuning ASR (FT. ASR) are evaluted.}
\resizebox{1.0\linewidth}{!}{
\begin{tabular}{c|c|cc|cc|cc} 
\toprule
\multicolumn{1}{c}{\multirow{2}{*}{Enhancement}} & \multicolumn{1}{c}{\multirow{2}{*}{Feature}} & \multirow{2}{*}{FT. SE} & \multirow{2}{*}{FT. ASR} & \multicolumn{2}{c|}{Dev. Set} & \multicolumn{2}{c}{Test Set}  \\ 
\cmidrule{5-8}
\multicolumn{1}{c}{}                             & \multicolumn{1}{c}{}                         &                          &                          & Simu. & Real                  & Simu. & Real                  \\ 
\midrule
\multirow{8}{*}{Conv-TasNet}                     & Fbank                                        & \xmark                        & \xmark                        & 17.22 & 16.76 & 30.28 & 32.50 \\
                                                 & Fbank                                        & \xmark                        & \checkmark                        & 11.42 & 9.92 & 21.16 & 21.82 \\
                                                 & Fbank                                       & \checkmark                        & \xmark                        & 9.20 & 8.33 & 17.01 & 16.56 \\
                                                 & Fbank                                        & \checkmark                        & \checkmark                        & 9.52 & 7.94 & 17.42 & 15.24 \\ 
\cmidrule{2-8}
                                                 & WavLM                                        & \xmark                        & \xmark                        & 5.96 & 4.37 & 13.52 & 12.11 \\
                                                 & WavLM                                        & \xmark                        & \checkmark                        & 5.45 & 4.04 & 12.68 & 11.57 \\
                                                 & WavLM                                        & \checkmark                        & \xmark                        & 3.54 & 2.27 & 6.73 & 4.90 \\
                                                 & WavLM                                        & \checkmark                        & \checkmark                        & \textbf{3.16} & \textbf{2.03} & \textbf{6.12} & \textbf{3.92} \\
\bottomrule
\end{tabular}
}
\label{tab:se-asr}
\end{table}
\vspace{-8pt}

\begin{table}[htb]
\centering
\vspace{-5pt}
\sisetup{table-format=2.1,round-mode=places,round-precision=1,table-number-alignment = center,detect-weight=true}
\caption{ASR performance (\%WER) comparison between the proposed IRIS model and the best existing single- and multi-channel systems.}
\vspace{-5pt}
\resizebox{0.9\linewidth}{!}{
\begin{tabular}{lc|cc|cc} 
\toprule
\multirow{2}{*}{System} & \multirow{2}{*}{Track}  & \multicolumn{2}{c|}{Dev. Set} & \multicolumn{2}{c}{Test Set}  \\ 
\cmidrule{3-6}
&  &  Simu.         & Real          & Simu.        & Real           \\ 
\midrule
IRIS (proposed) & 1ch & 3.16 & 2.03 & 6.12 & 3.92 \\
Yang \textit{et al.} \cite{yang2022conformer} & 1ch & 4.99 & 3.35 & 8.61         & 6.25  \\
\midrule
Du \textit{et al. \cite{du2016ustc}} & 2ch & 3.46 & 2.33 & 5.74 & 3.91           \\
Wang \textit{et al. \cite{wang2020complex}} & 2ch & 2.17 & 1.99 & 2.53 & 3.19           \\
\midrule
Kaldi Baseline \cite{chen2018building} & 6ch & 1.90 & 2.10 & 2.74 & 2.66 \\
Wang \textit{et al. \cite{wang2020complex}} & 6ch & 1.15 & 1.50 & 1.45 & 1.99           \\
\bottomrule
\end{tabular}
}
\label{tab:comparison}
\end{table}
\vspace{-8pt}

Next, we evaluate our proposed IRIS models. From the results in Table~\ref{tab:e2e-asr-results}, we already know that WavLM is robust in the noisy condition. In this part, we further investigate if adding a speech enhancement module is beneficial to the model. As a reference, we did the similar evaluation on the E2E-ASR based on Fbank. Considering the computation cost when concatenated with the ASR model, we choose Conv-TasNet as the enhancement model and reduce the number of parameters by using a shallow architecture described in Sec.~\ref{ssec:config}. The SI-SNRs of the pre-trained speech enhancement model are $9.55$ dB and $9.71$ dB on the development and test sets, respectively.

First, we directly concatenate the speech enhancement model and E2E-ASR models to perform the speech recognition. The results are shown in the Table~\ref{tab:se-asr}. If the simple concatenation is used, both the performance of the Fbank-based system and that of the WavLM-based system are degraded, compared with the results of system 5 and 7 in the previous table. This indicates that speech enhancement models do not necessarily improve the ASR performance on noisy speech, because the training objectives of speech enhancement and recognition are not very well aligned. It is a well-known phonomenon in previous research \cite{Closing-Zhang2021}. 

Second, if we keep the enhancement model fixed and fine-tune the ASR model with ASR loss, the performance of a WavLM-based system is slightly improved but not reaching the same level as system 7 in the previous table. We believe the artifacts from the enhancement model is difficult to handle by the WavLM. For the Fbank-based system, the performance degradation is mitigated. However, in the other way around, if we keep the ASR model fixed and fine-tune the enhancement model with both enhancement loss and the ASR loss, we find that the performance are significantly improved in WavLM-based model and Fbank-based model, especially on the simulation sets. We assume that the major reason is because only the simulation data is used to fine-tune the enhancement module.

As the last case, we fine-tune both enhancement and ASR models with the enhancement and ASR losses. We observe further improvements on both Fbank-based and WavLM-based models. For the WavLM system, the best performance is achieved. Compared to the system 7 in the previous table, WavLM without speech enhancement, WERs on all the evaluation sets are further improved with a nonneglectable improvement. In Table~\ref{tab:comparison}, we list the best result of existing systems from Table~\ref{tab:e2e-asr-results}, the result of the 1st ranking system in CHiME-4 two- and six-channel track \cite{du2016ustc} and the result of our end-to-end IRIS system. Our system achieves a new state-of-the-art performance on the monaural CHiME-4 ASR task\footnote{The pre-trained SSLR has more parameters and uses more data.}, outperforming the best monaural system. More interestingly, the results are comparable to the CHiME-4  challenge best 2-channel results from \cite{du2016ustc}.

The results indicate that the noise robust SSLR can still suffer from the degradation of noise. We can greatly alleviate the problem by introducing a speech enhancement as pre-processing. However, it is critical to fine-tune both models jointly to eliminate the mismatch. This rule can be applied to Fbank-based E2E-ASR model as well. 

\vspace{-3pt}
\subsection{Analysis}
\label{ssec:analysis}
\vspace{-3pt}

\begin{figure}[htb]
    \vspace{-6pt}
    \centering
    \includegraphics[width=\linewidth]{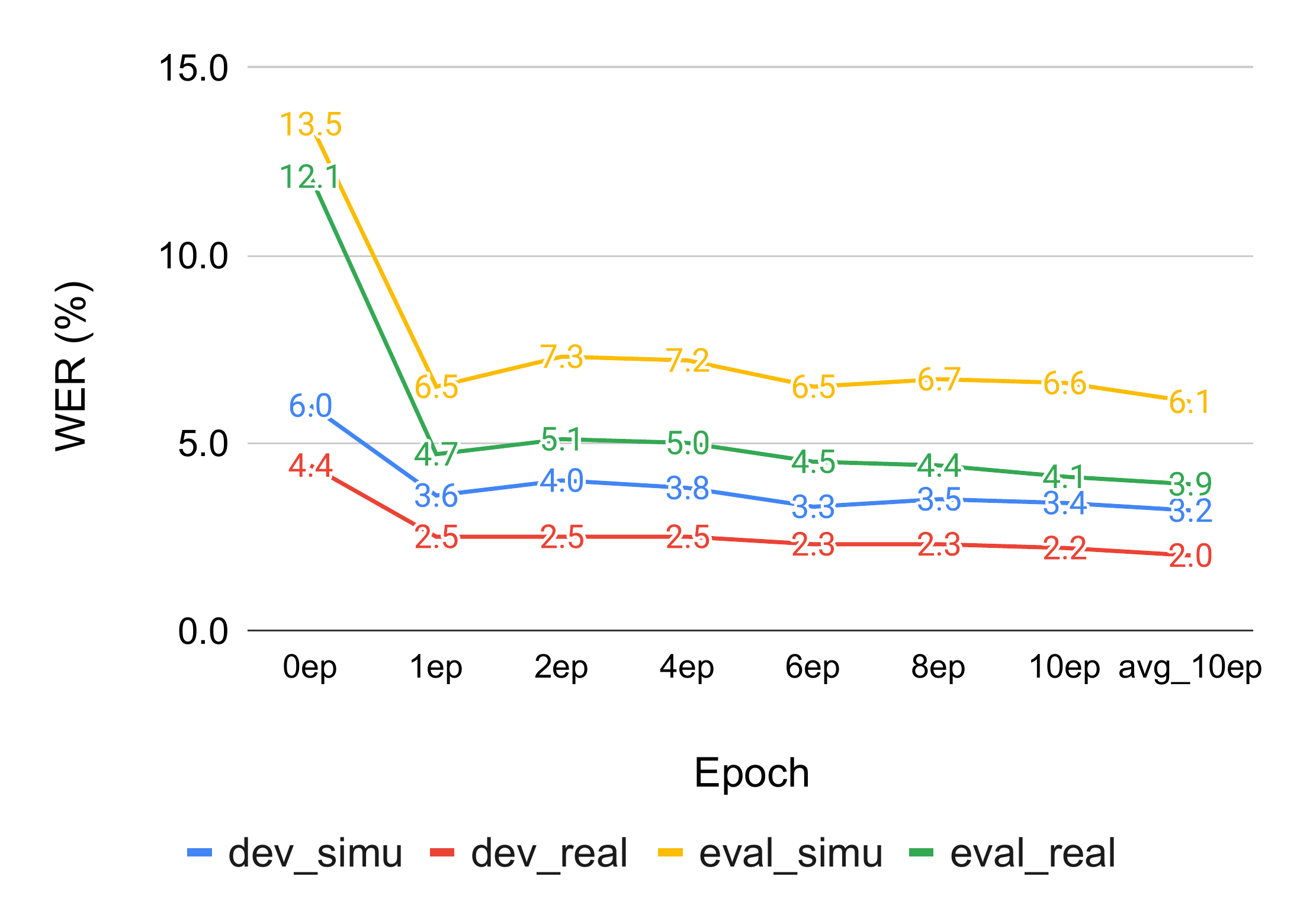}
    \caption{CHiME-4 ASR performance (WERs) of the IRIS model at different epochs during fine-tuning. Both SE and ASR are fine-tuned.}
    \label{fig:chart}
    \vspace{-8pt}
\end{figure}

It is interesting to know how the fine-tuning improves the IRIS model. We show the ASR performance with the checkpoints in the middle of fine-tuning the IRIS model in Figure.~\ref{fig:chart}. It can be observed that the fine-tuning converges very fast. After only one epoch, the WERs can reach a very good level. With the model average over the first 10 epochs, the best performance can be observed.

\begin{figure}[htb]
    \vspace{-6pt}
    \centering
    \includegraphics[width=0.9\linewidth]{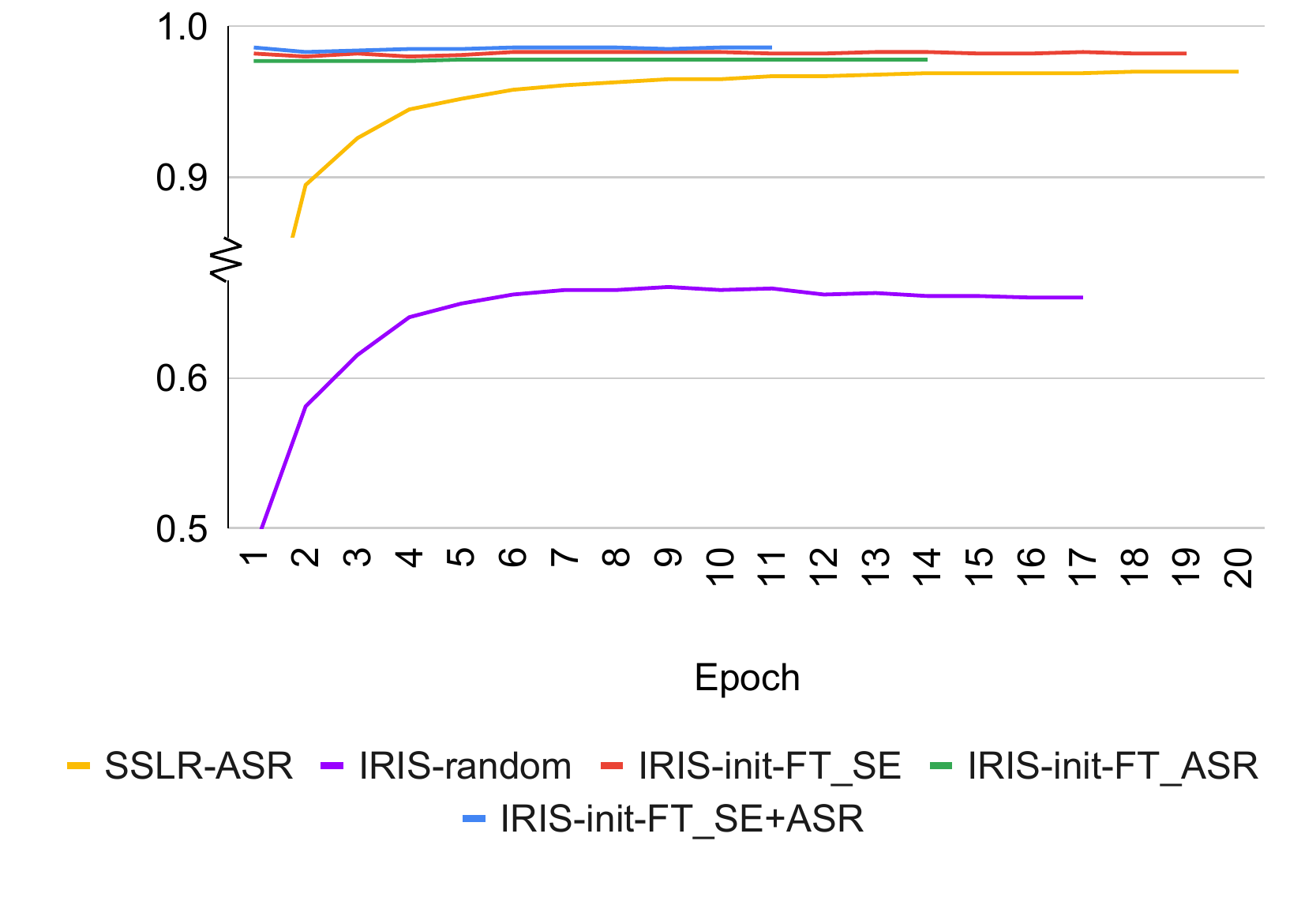}
    \caption{Accuracies on the development set for training and fine-tuning different models.}
    \label{fig:training_curve}
    \vspace{-6pt}
\end{figure}

One difficult point is that the current IRIS model needs pre-training, taking extra efforts to prepare the individual enhancement and ASR models. In Figure~\ref{fig:training_curve}, we show the training curves of the following models:
{\footnotesize
\begin{align}
\begin{array}{lcr}
\toprule
  \text{Model}   &  \text{Init. Param.} & \text{Update Param.} \\
  \midrule
  \text{SSLR-ASR}  &  \hat{\theta}^{sslr}  & \theta^{asr} \\
  \text{IRIS-random}  &  \hat{\theta}^{sslr}  & \theta^{se}, \theta^{asr} \\
  \text{IRIS-init-FT\_SE}  & \hat{\theta}^{se}, \hat{\theta}^{sslr}, \hat{\theta}^{asr}  & \theta^{se} \\
  \text{IRIS-init-FT\_ASR}  & \hat{\theta}^{se}, \hat{\theta}^{sslr}, \hat{\theta}^{asr}  & \theta^{asr} \\
  \text{IRIS-init-FT\_SE+ASR}  & \hat{\theta}^{se}, \hat{\theta}^{sslr}, \hat{\theta}^{asr}  & \theta^{se}, \theta^{asr} \\
\bottomrule
\end{array} \nonumber
\end{align}
}

We can see that training the IRIS model from random initialization could not converge to a good point. We assume that the deep architecture of the SSLR models might disturb the gradient back-propagation from ASR to the enhancement. More training tricks are required. However, if we initialize the parameters of each module, the training reaches a good level after the 1st epoch.

\vspace{-4pt}
\section{Conclusions}
\label{sec:conclusion}
\vspace{-4pt}
We have proposed a new end-to-end model, IRIS, for robust speech recognition. The model contains three modules including an SE module, an SSLR module and an ASR module. For the implementation, we use Conv-TasNet as SE module, WavLM as SSLR module and a joint CTC/attention-based encoder-decoder as ASR module. In the evaluation on monaural CHiME-4 task, the IRIS model outperforms the current state-of-the-art system, which is based on the hybrid ASR model. It should be noted that the pre-training of SSLR model uses more data and more parameters. Moving forward, we plan to explore the generalization ability of our proposed system and to investigate the way to enable the end-to-end joint training from random initialization.

\vspace{-4pt}
\section{Acknowledgements}
\label{sec:ack}
\vspace{-4pt}
This work used the Extreme Science and Engineering Discovery Environment (XSEDE) \cite{towns2014xsede}, which is supported by National Science Foundation grant number ACI-1548562. Specifically, it used the Bridges system \cite{nystrom2015bridges}, which is supported by NSF award number ACI-1445606, at the Pittsburgh Supercomputing Center (PSC).

\bibliographystyle{IEEEtran}

\bibliography{mybib}


\end{document}